# Planar and 3-dimensional damage free etching of $\beta$-Ga$_2$O$_3$ using atomic gallium flux

Nidhin Kurian Kalarickal[1], Andreas Fiedler[1], Sushovan Dhara[1], Mohammad Wahidur Rahman[1], Taeyoung Kim[1], Zhanbo Xia[1], Zane Jamal Eddine[1], Ashok Dheenan[1], Mark Brenner[1], Siddharth Rajan[1,2]

[1]*Department of Electrical and Computer Engineering, Ohio State University*

[2] *Department of Materials Science and Engineering, Ohio State University*

In-situ etching using Ga flux in an ultra-high vacuum environment like MBE is introduced as a method to make high aspect ratio 3 dimensional structures in $\beta$-Ga$_2$O$_3$. Etching of $\beta$-Ga$_2$O$_3$ due to excess Ga adatoms on the epilayer surface had been viewed as non-ideal for epitaxial growth especially since it results in plateauing and lowering of growth rate. In this study, we use this well-known reaction from epitaxial growth of $\beta$-Ga$_2$O$_3$, to intentionally etch $\beta$-Ga$_2$O$_3$. We demonstrate etch rate ranging from 2.9 nm/min to 30 nm/min with the highest reported etch rate being only limited by the highest Ga flux used. Patterned in-situ etching was also used to study the effect of fin orientation on the sidewall profiles and dopant (Si) segregation on the etched surface. Using in-situ Ga etching, we also demonstrate 150 nm wide fins and 200 nm wide nano pillars with high aspect ratio. This new etching method could enable future development of highly scaled vertical and lateral 3D devices in $\beta$-Ga$_2$O$_3$.

The ultra-wide band gap semiconductor, $\beta$-Ga$_2$O$_3$ has attracted a lot of interest owing to its large breakdown field strength of 8 MV/cm [1–3]. When compared to existing state of the art technologies like Si, SiC and GaN, the high breakdown field strength theoretically predicts better performance for $\beta$-Ga$_2$O$_3$ based devices especially for applications like high voltage switching and high frequency power amplification [3]. In addition, the availability of bulk substrates grown from melt based techniques [4–7] and the wide range of controllable doping ($10^{15}$ cm$^{-3}$ to $10^{20}$ cm$^{-3}$) [8,9] has led to rapid development of $\beta$-Ga$_2$O$_3$ devices in both vertical and lateral topology with excellent performance [10–13].

Due to the lack of p-type doping, most vertical devices in $\beta$-Ga$_2$O$_3$ require confined and scaled regions that control the flow of current between the source and drain, like vertical fins and pillars [10,14]. Using vertical fin structures also improves the electric field distribution in vertical Schottky barrier diodes by reducing the electric field seen at the Schottky metal semiconductor interface [15,16]. In addition, moving to a fin geometry would also result in increased power density and possibly enhancement-mode operation in lateral devices. Fabrication of these 3-dimensional structures require controlled, damage free etching that ideally provides vertical sidewalls (90º sidewall angle). Most dry etching recipes for etching $\beta$-Ga$_2$O$_3$ are based on chlorine and argon, and have been found to cause significant etch damage resulting in non-ideal device characteristics [17,18]. Wet etching recipes have also been demonstrated using HF, H$_3$PO$_4$ (hot) and KOH (hot) [19–21], but wet etching generally provides slanted sidewalls and poorly controlled etch depths which are not ideal for fabricating scaled submicron fins. In addition to traditional dry and wet etching techniques, metal assisted chemical etching (MacEtch) for $\beta$-Ga$_2$O$_3$ was also demonstrated to produce 3D fin structures [22,23]. However, the MacEtch process was found to result in slanted sidewalls (except for fins oriented perpendicular to [102] direction) along with reduced Schottky barrier heights on etched sidewalls [22].

In $\beta$-Ga$_2$O$_3$, the epitaxial growth proceeds via competition between the sesquioxide (Ga$_2$O$_3$) and the suboxide phases (Ga$_2$O). The suboxide being a volatile phase, its formation and subsequent desorption from the epilayer surface results in the reduction of Ga adatoms that gets incorporated into the Ga$_2$O$_3$ epilayer resulting in the reduction of growth rate [24]. Additionally, in the absence of active oxygen, exposure of $Ga_2O_3$ surface to Ga flux results in etching of the epilayer according to [24]

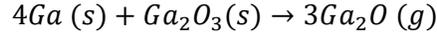

$$4Ga\ (s) + Ga_2O_3(s) \rightarrow 3Ga_2O\ (g)$$

In this work, we intentionally use the etching reaction given above as an in-situ damage free etch method for $Ga_2O_3$. Etch rates ranging from 2.9 nm/min to 30 nm/min are demonstrated with highest etch rate only limited by the supplied Ga flux. We also show successful patterned etching of $\beta$-$Ga_2O_3$ using the same method to produce high aspect ratio 3D structures like fins, trenches and nano pillars with vertical sidewalls. The effect of Ga etching on dopant segregation on the etched surface is also examined by fabricating Schottky diodes on the etched surface.

Ga etching reported throughout this work was performed in the ultra-high vacuum environment of a Veeco Gen 930 molecular beam epitaxy (MBE) system. The base pressure in the MBE chamber was found to be roughly around $5 \times 10^{-10}$ Torr over the course of this study. An Escience Titan cell [25] was used as the Ga Knudsen cell with Ga flux ranging from $1.5 \times 10^{-7}$ Torr to $1.3 \times 10^{-6}$ Torr used in this study. The Ga flux was measured using a beam flux monitor placed at the substrate position. Three different orientations of $\beta$-$Ga_2O_3$ samples were used, namely (010), (001) and ($\bar{2}$01). Fe doped substrate [26], Sn doped substrate[26] and HVPE grown epilayer[26] (6 $\mu$m epi, $N_D$-$N_A$ = $6 \times 10^{16}$ cm$^{-3}$) were used for the (010), ($\bar{2}$01) and (001) orientations respectively. The samples were bonded to Silicon carrier wafers using Indium and was heated to the required etch temperature. Etch temperatures ($T_{sub}$=thermocouple temperature) used in this study are 800 °C, 700 °C and 550 °C.

Fig.1 (a) and (b) show the measured etch rate as a function of Ga flux for the three different substrate orientations at $T_{sub}$=800 °C and 700 °C. Etch rates for $T_{sub}$=550 °C is not provided due to significant roughening of the surface (discussed later). The samples were patterned using $SiO_2$ hard mask (deposited using PECVD, ~100 nm thick) and spaces were opened where the $\beta$-$Ga_2O_3$ surface was exposed. The etch rate was then measured using atomic force microscopy (AFM) by measuring the etch depth obtained after Ga etching. The etch rate increases linearly from ~2.9 nm/min to ~30 nm/min as the Ga flux is increased from $1.5 \times 10^{-7}$ Torr to $1.3 \times 10^{-6}$ Torr. Etch rates below ~ 10 nm/min are ideal for processes like gate recess and regrowth etching whereas high etch rates around 30 nm/min would be useful for deep etching to fabricate 3 dimensional structures like fins and trenches. Note that etch rates above 30 nm/min are possible and would only require an increase in Ga flux. No significant influence of substrate temperature is observed on etch rate when comparing $T_{sub}$=800 °C and $T_{sub}$=700 °C. This would suggest that thermal desorption of Ga from the β-$Ga_2O_3$ surface does not contribute significantly to loss of Ga ad atoms at the high Ga fluxes used in this study. Additionally, each substrate orientation is found to show similar etch rate at a given Ga flux (Fig.1 (a) and (b)). This is expected since the etch rate based on equation (1) would only depend on the number density of β-$Ga_2O_3$ (number of β-$Ga_2O_3$ molecules/unit cell volume) which is independent of orientation. Therefore, the wide temperature range and similar etch rates on all substrate orientation provides improved flexibility and makes this method an excellent choice for etching β-$Ga_2O_3$.

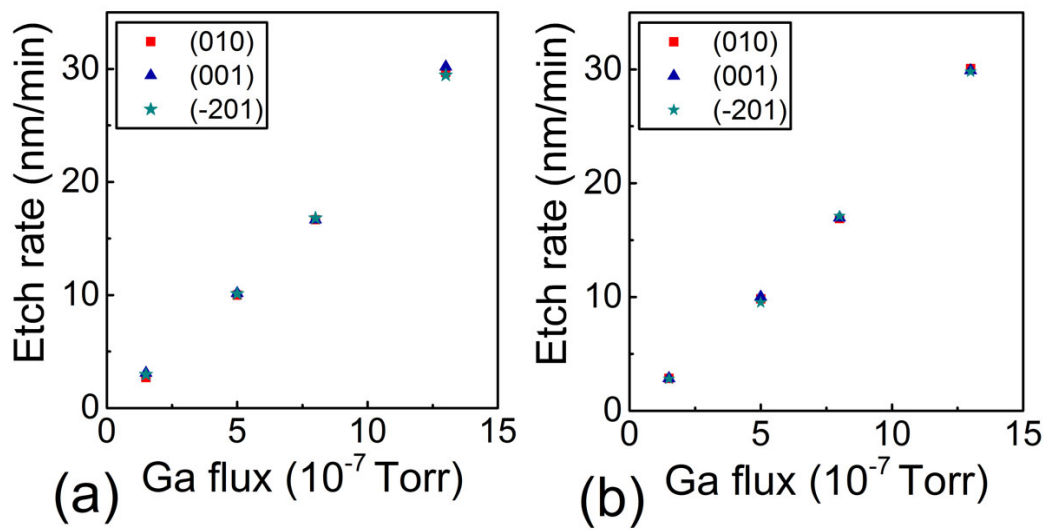

Fig.1 (a) Etch rate as a function of Ga flux at $T_{sub}$=800 °C. (b) Etch rate as a function of Ga flux at $T_{sub}$=700 °C.

The surface morphology (AFM) of the etched surfaces are shown in Fig.2 for the three different sample orientations. The AFM measurements were carried out on samples which were etched at a Ga flux of $8 \times 10^{-7}$ Torr for 15 mins resulting in an etch depth of ~300 nm. Smooth surfaces are obtained at $T_{sub}$=800 °C and $T_{sub}$=700 °C whereas significant roughening is obtained at $T_{sub}$=550 °C. $T_{sub}$=700 °C was found to give the smoothest surface morphology with surface roughness of ~ 2.2 nm or less as shown in Fig.2. For etching at $T_{sub}$=550 °C, formation of deep trenches (for (010) orientation) and pits (for (001) and ($\bar{2}$01) orientation) are evident from the AFM scans which results in surface roughening. Optical microscopy of the samples post Ga etching revealed absence of Ga droplets on the β-$Ga_2O_3$ surface after etching at $T_{sub}$=800 °C and $T_{sub}$=700 °C, while Ga droplet accumulation was observed at $T_{sub}$=550 °C. However, Ga droplets were observed on the $SiO_2$ hard mask regardless of the $T_{sub}$ used for etching. At low Ga flux (~$1.5 \times 10^{-7}$ Torr), smooth etched surfaces without the formation of pits or trenches were also obtained at $T_{sub}$=550 °C (Fig.4 (c)). This would suggest that etching at lower substrate temperatures ($T_{sub}$=550 °C) is limited to low Ga fluxes above which the low ad atom mobility results in non-uniform etching profiles.

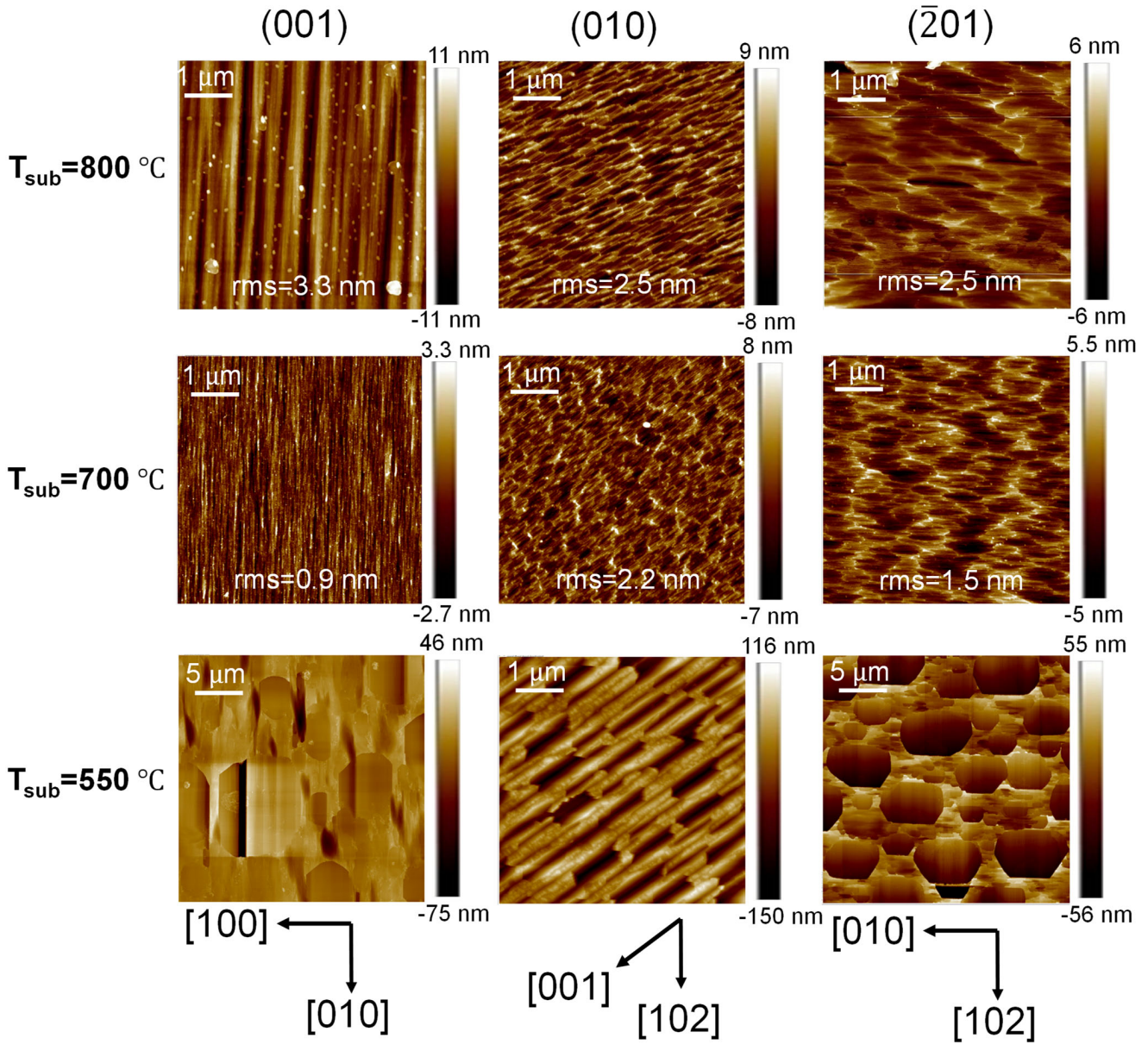

Fig.2 AFM images of Ga etched surface at $T_{sub}$=800 °C, $T_{sub}$=700 °C and $T_{sub}$=550 °C at $\phi_{Ga}$=8x10$^{-7}$ Torr for three different $\beta$-Ga$_2$O$_3$ substrate orientation. Larger (25 by 25 $\mu$m) scans are shown for (001) and ($\bar{2}$01) samples etched at $T_{sub}$=550 °C to clearly show the etch pits.

Deep etching of patterned surfaces was also studied for making 3D structures. Only the (010) and (001) orientations were considered for this study since most of the $\beta$-Ga$_2$O$_3$ epilayers are currently grown along these two orientations. We used spoke wheel structures as shown in Fig.3 (a) and (b) for deep etching to study the effect of fin orientation on the sidewall profile. The spoke wheel structures were patterned using $SiO_2$ as previously mentioned, with each fin having a width of 3 $\mu$m. The fins are labelled according to the sidewall plane (hkl) and important in-plane crystallographic directions are provided for reference as shown in Fig.3 (a) and (b). The samples were etched at a high Ga flux of $8 \times 10^{-7}$ Torr at $T_{sub}$=700 °C for a total time of 2 hours resulting in an etch depth of ~ 2 $\mu$m. $T_{sub}$=700 °C was chosen as it gave the best surface morphology as explained before. The etched samples were then observed under scanning electron microscope (SEM) with the sample tilted. The obtained sidewall profiles on both (010) and (001) oriented samples are shown in Fig.3 (c), (d) and in the supplementary material (SFig.1 (b) and SFig.2 (b)). Surprisingly, vertical sidewall profiles are obtained for all orientations. This is somewhat unexpected as the Ga cell is not mounted vertically with respect to the substrate but is at an angle of 51° with respect to the plane of the substrate. Additionally, comparison of the fin width before and after Ga etching showed that in addition to the vertical etching of the $\beta$-$Ga_2O_3$ surface, the fin sidewalls were also etched inward resulting in a reduction of the fin width. Therefore, the Ga etch mechanism for deep etching may be schematically illustrated as shown in Fig.3 (e). The etching process starts with $\beta$-$Ga_2O_3$ surface patterned using $SiO_2$ which is then loaded into the MBE chamber and heated to $T_{sub}$. Upon exposure to Ga flux, the vertical $\beta$-$Ga_2O_3$ surface is etched down at a rate of $\lambda$ (nm/min) which depends on the supplied Ga flux. The sidewall is also etched inward maintaining a vertical sidewall profile at a rate $\zeta\lambda$ (nm/min), where $\zeta$ is termed as the sidewall etch factor and is a number between 0 and 1. The value of $\zeta$ was found to be around 0.2 at $T_{sub}$=700 °C, meaning that after a vertical etch of 2 $\mu$m, a 3 $\mu$m wide fin is reduced to a width of ~ 2.2 $\mu$m due to etching of the sidewalls by 0.4 $\mu$m from each side. The formation of a vertical sidewall suggests that Ga adatoms are able to diffuse freely resulting in a uniform concentration of Ga ad atoms on the sidewall even in areas that are in the shadow region of the flux. In addition, the accumulated Ga on top of the $SiO_2$ mask may also be acting as a source of Ga ad atoms especially to the shadow region (Fig.3 (f)). Therefore, the sidewall etch factor $\zeta$ in general is a function of the angle to the Ga cell, the substrate temperature and the fin orientation. A more detailed analysis on the sidewall etch factor and its mechanism is currently outside the scope of this work.

As shown in Fig.3 (c), (d) and supplementary material (SFig.1 (b) and SFig.2 (b)), both rough and smooth sidewalls are obtained depending on the fin orientation. For the (010) orientated sample (Fig.3 (c) and SFig.1 (b)), fins with sidewall planes of (001), (100), (101) and ($\bar{1}$02) are found to give the smoothest sidewalls while other fin orientations result in rough sidewalls. Surprisingly, this does not fully match with the lowest energy surfaces[27] since ($\bar{2}$01) and ($\bar{1}$01) do not show smooth surfaces. Previous studies on wet etching of $\beta$-Ga$_2$O$_3$ ((010) oriented samples) showed that fins that are oriented along the [001] direction (forms (100) sidewall planes) form the smoothest sidewalls. In case of Ga etching, the (100) sidewalls are found to form steps that are separated by several microns (Fig.3 (c)) which makes them smooth on a small length scale and rough on a large length scale. For the (001) oriented sample, (100) and (210) sidewalls are found to be the smoothest. As mentioned before, due to the lateral etching involved, fin widths are reduced as a function of time. Utilizing this we achieved submicron structures with high aspect ratios (height/width) as shown in Fig.3 (f), (g) and supplementary material (SFig.2 (c), (d)). Using an initial fin width of ~1 $\mu$m in the SiO$_2$ mask, Ga etching for 2 hours at a flux of 8x10$^{-7}$ Torr and T$_{sub}$=700 °C resulted in fin widths as low as ~ 150 nm with fin height of ~1.5 $\mu$m on (010) oriented samples (Fig.3 (f)). In addition to submicron fins, vertical $\beta$-Ga$_2$O$_3$ nano pillars with diameter of ~200 nm are also demonstrated (Fig.3 (f)) using the same method. The aspect ratios of fins and nano pillars obtained in this work are extremely high and could not be obtained without the lateral sidewall etch mechanism that maintains a vertical sidewall profile.

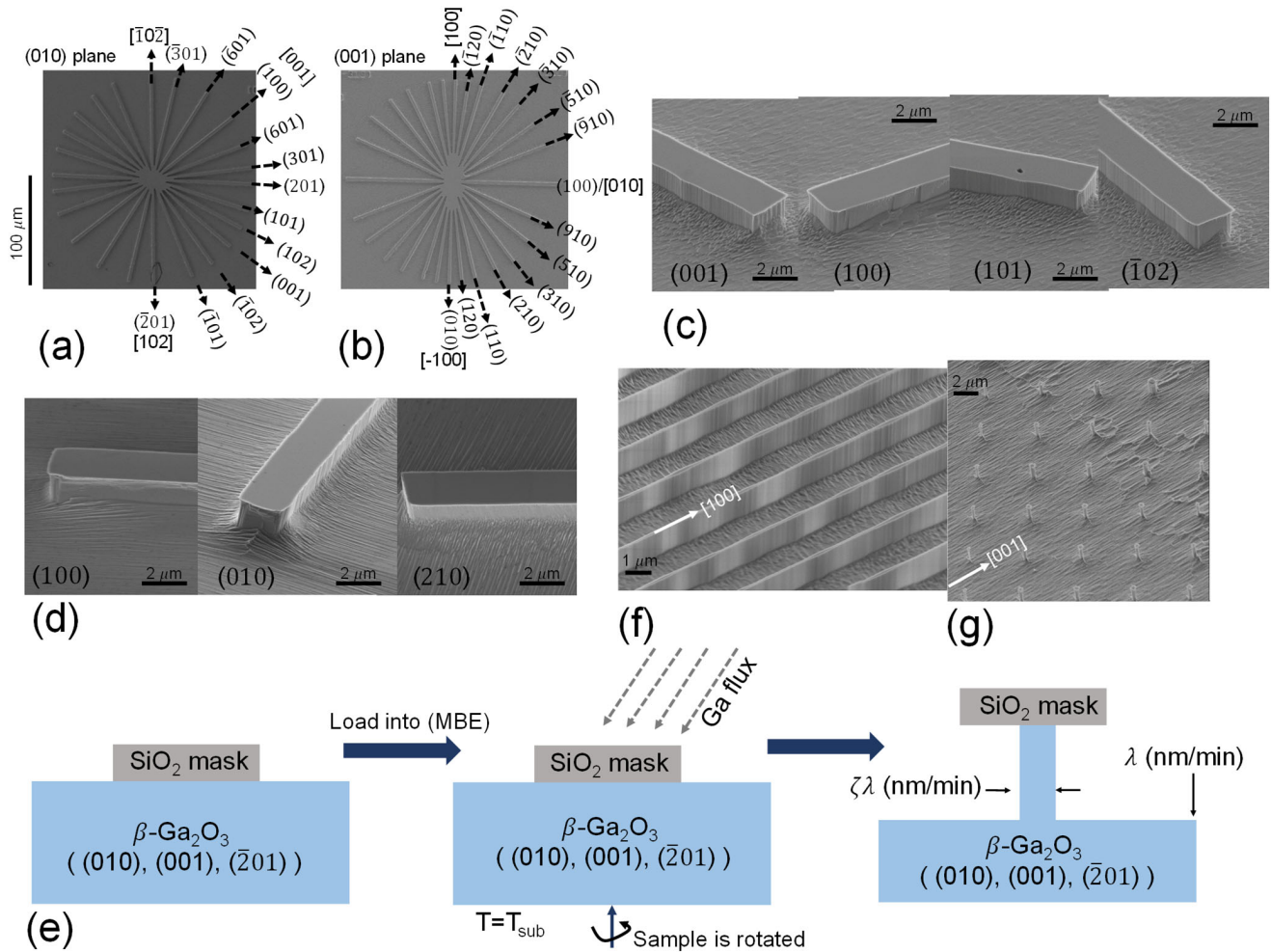

Fig.3 (a) SEM image of spoke wheel pattern used on (010) sample. Each fin is labelled using the sidewall plane, [102] and [001] in-plane directions are provided for clarity. (b) SEM image of spoke wheel pattern used on (001) sample. [100] and [010] directions are provided for clarity. (c) Sidewall profiles of (001), (100), (101) and ($\bar{1}$02) planes obtained on (010) sample at $T_{sub}$=700 °C. Remaining sidewall profiles are provided in SFig.1 (b) (supplementary material) (d) Sidewall profiles of (100), (010) and (201) planes obtained on (001) sample at $T_{sub}$=700 °C. Remaining sidewall profiles are provided in SFig.2 (b) (supplementary material) (e) SEM image of ~150 nm wide fins obtained on (010) sample. (f) 200 nm diameter nano-pillars obtained on (010) sample.

In addition to quantifying the etching characteristics, it is also important to study the surface segregation effects of this etching method. Unlike Ga$_2$O, many of the dopants and deep acceptors that are present in $\beta$-Ga$_2$O$_3$ epilayers and substrates are not volatile. This could result in segregation of these elements on the etched $\beta$-Ga$_2$O$_3$ surface resulting in increased surface concentrations. We investigated the segregation of Si after Ga etching by using an HVPE grown, (001) oriented epilayer with a Si dopant concentration of 6x10$^{16}$ cm$^{-3}$ (N$_D$-N$_A$). These samples were patterned using SiO$_2$ and then etched using Ga. After the etch process, the SiO$_2$ was removed using buffered oxide etch (BOE) and circular Schottky contacts (Ni/Au – 30/100 nm) were patterned and deposited on the etched and unetched surfaces. Capacitance voltage (C-V) and I-V characteristics were measured and compared to study the quality of the etched surface as well as dopant segregation. Fig.4 (a) shows the extracted charge density (from C-V) on the etched and unetched surfaces after Ga etching at $T_{sub}$=700 °C for 2 hours (vertical etch depth of 2 $\mu$m). Clear segregation of Si is observed in the top layer of the etched surface. This increased surface concentration is also visible in the schottky I-V characteristics as severe leakage in reverse bias (Fig.4 (b)). The etched surface also shows a higher ideality factor ($\eta$) of 1.3 compared to the unetched surface ($\eta$=1.04). The increased surface concentration of Si is likely due to the diffusion of Si atoms that are left behind from the etching process into the top layer of the etched surface. This increased surface concentration may be useful in some applications like contact regrowth or field emission, but in general is detrimental as evidenced by the high reverse leakage current in Fig. 4 (b). In comparison, etching at $T_{sub}$=550 °C ($\phi_{Ga}$=8x10$^{-7}$ Torr, etch depth=500 nm), does not show any surface segregation of Si as shown in Fig.4 (b). The reverse leakage in Schottky diodes fabricated on the etched surface also shows no increase when compared to those on unetched surface. A small increase in the ideality factor to 1.05 is observed when compared to devices on unetched surface ($\eta$=1.02), which is likely due to the presence of pits on the $\beta$-Ga$_2$O$_3$ surface etched at

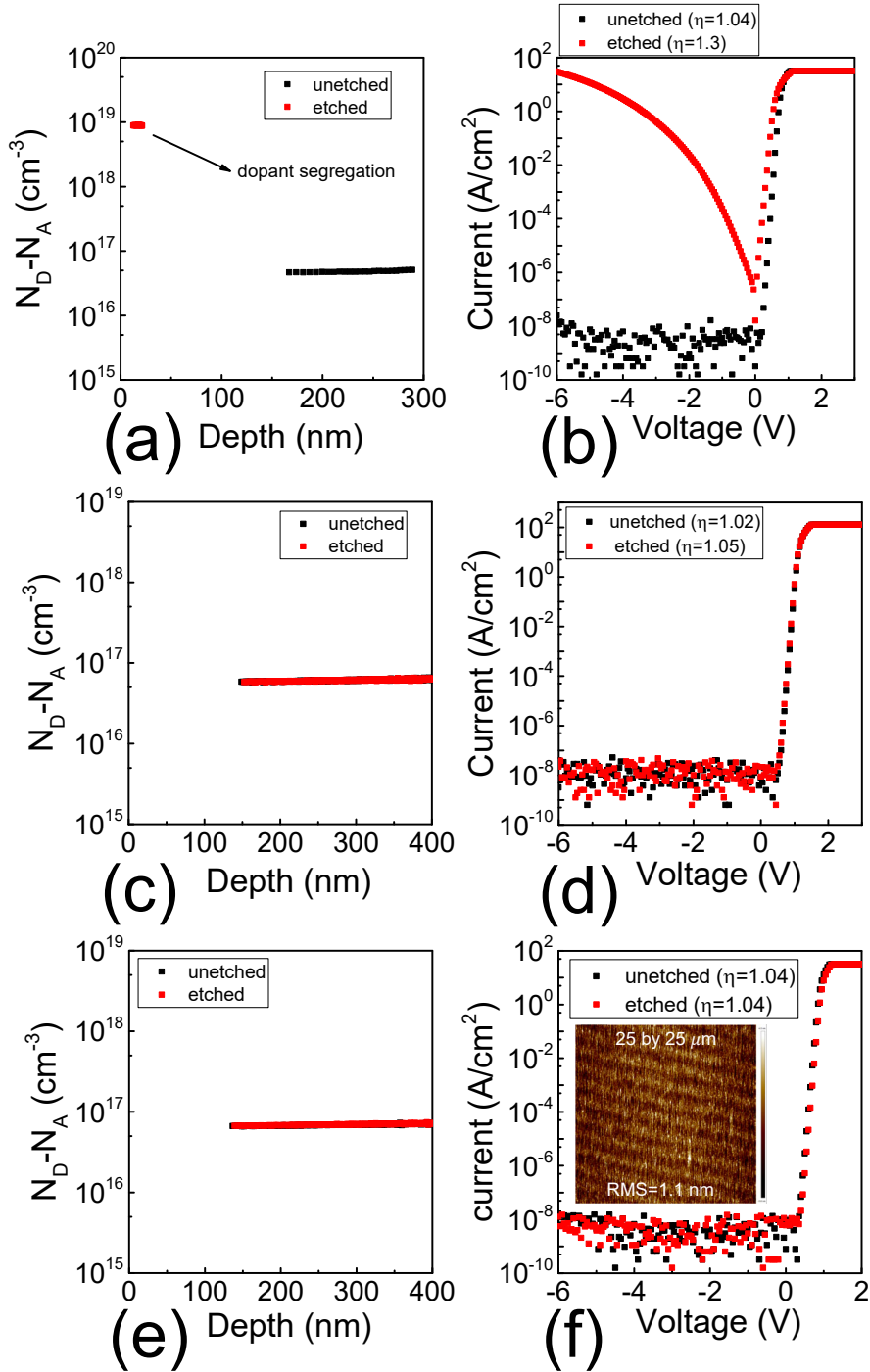

Fig.4 (a) Charge profile extracted on Ga etched surface at $T_{sub}$=700 °C. (b) I-V characteristics of Schottky diodes fabricated on Ga etched surface at $T_{sub}$=700 °C. (c) Charge profile extracted on Ga etched surface at $T_{sub}$=550 °C. (d) I-V characteristics of Schottky diodes fabricated on Ga etched surface at $T_{sub}$=550 °C. (e) Charge profile extracted on Ga etched surface after two-step etch ($T_{sub}$=700 °C and $T_{sub}$=550 °C) (f) I-V characteristics of Schottky diodes fabricated on Ga etched surface after two-step etch. Inset shows the AFM scan (25 by 25 $\mu$m) of $\beta$-Ga$_2$O$_3$ (001) surface after two-step etching. The ideality factor given is the average value over 6 orders increase in current.

$T_{sub}$=550 °C at high Ga flux. To solve this problem, we introduced a two-step etch process where at first the required etch depth is achieved at $T_{sub}$=700 °C ($\phi_{Ga}$=8x10$^{-7}$ Torr), following which a second etch step is carried out at lower Ga flux ($\phi_{Ga}$=1.5x10$^{-7}$ Torr) to remove the surface segregation layer (<100 nm) at $T_{sub}$=550 °C. Fig.4 (e) shows the extracted charge density after the two-step etch method showing successful removal of the top surface segregation layer. The lower etching temperature likely prohibits the diffusion of Si back into the top layer of the etched surface limiting dopant segregation. The inset in Fig.4 (e) also shows the smooth surface morphology after the two-step etching. I-V characteristics of Schottky diodes fabricated on the etched surface (two-step etch) is also found to be identical to those on the unetched surface both in terms of the reverse leakage and the ideality factor ($\eta$=1.04).

In conclusion, we introduce a new etching method that is unique to $\beta$-Ga$_2$O$_3$ by using Ga as an etchant in a high vacuum environment. This method provides a low-damage, efficient method to etch $\beta$-Ga$_2$O$_3$ by controlling the supplied Ga flux. In addition to the vertical etch, a non-zero lateral etch rate of the sidewalls is also observed using this method which results in the reduction of feature size as a function of time but at the same time also maintains a vertical sidewall profile. Using this method, we successfully demonstrated 150 nm fins and 200 nm vertical nano pillars with high aspect ratio. A two-step etching scheme was also demonstrated to remove segregation of Si atoms from the top surface of the etched epilayer resulting in Schottky diodes with near unity ideality factors and low reverse leakage currents. This etching method could enable future development of a variety of scaled vertical and lateral devices in $\beta$-Ga$_2$O$_3$.

## ACKNOWLEDGEMENTS

This work was supported by the Air Force Office of Scientific Research under AFOSR Award FA9550-18-1-0479 and AFOSR Award FA9550-19-1-0349.

# Planar and 3-dimensional damage free etching of $\beta$-Ga$_2$O$_3$ using atomic gallium flux

Nidhin Kurian Kalarickal[1], Andreas Fiedler[1], Sushovan Dhara[1], Mohammad Wahidur Rahman[1], Taeyoung Kim[1], Zhanbo Xia[1], Zane Jamal Eddine[1], Ashok Dheenan[1], Mark Brenner[1], Siddharth Rajan[1,2]

[1]*Department of Electrical and Computer Engineering, Ohio State University*

[2] *Department of Materials Science and Engineering, Ohio State University*

This section contains two Figures, SFig.1 and SFig.2 which shows the sidewall profile of fins that are etched using in-situ Ga etching. SFig.1 (a) and (b) corresponds to fins formed on the (001) $\beta$-Ga$_2$O$_3$ sample (HVPE grown epilayer) and SFig.2 (a) and (b) corresponds to the fins formed on (010) $\beta$-Ga$_2$O$_3$ sample (Fe doped substrate). SFig.1 (c) and (d) also shows submicron fins (900 nm and 550 nm) fabricated on (001) $\beta$-Ga$_2$O$_3$ sample using the same method.

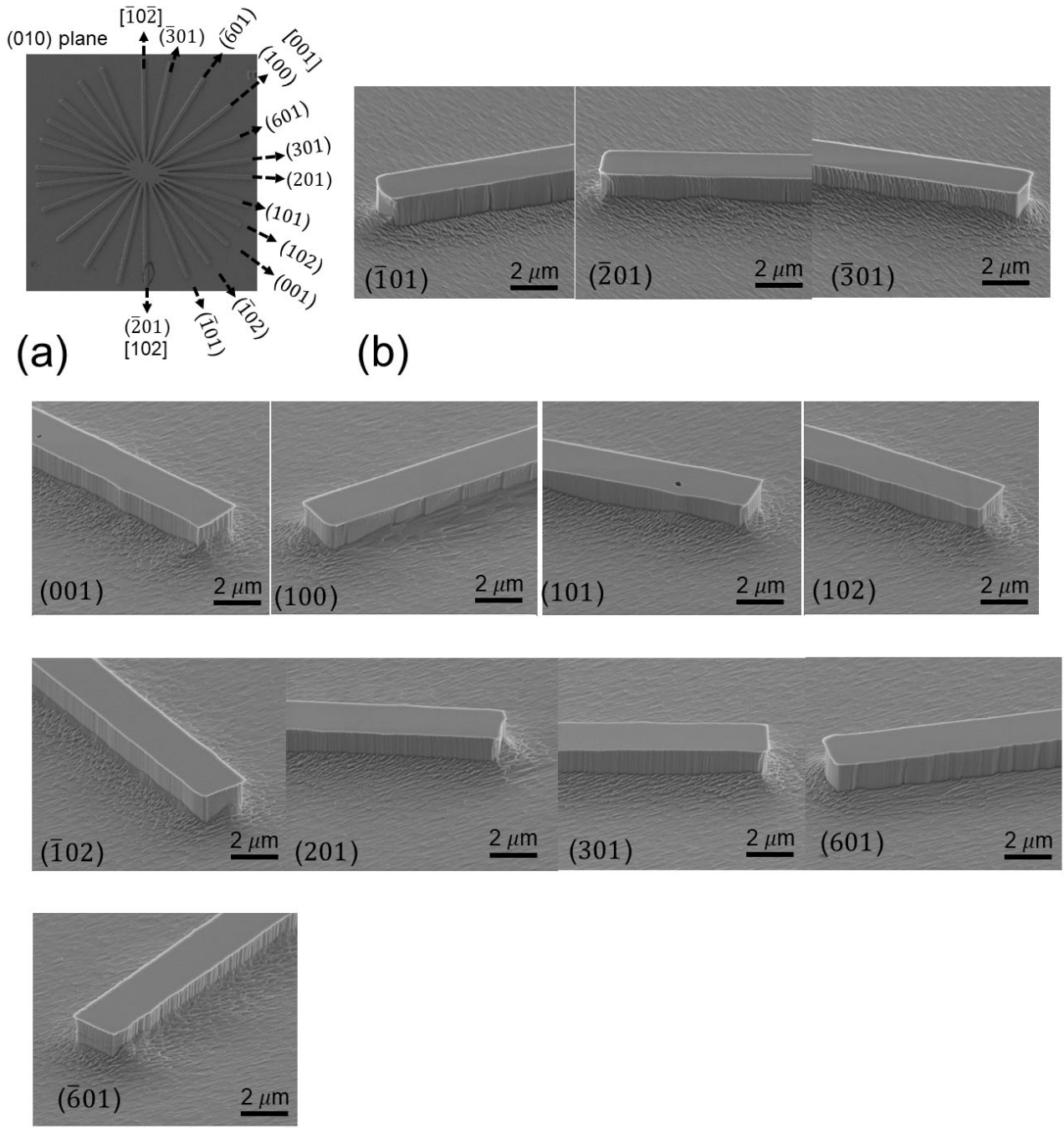

SFig.1 (a) Spoke wheel pattern used for the etching study. The fins are marked according to the sidewall planes. [102] and [001] directions are provided for clarity (b) SEM images of the sidewalls for different fin orientation.

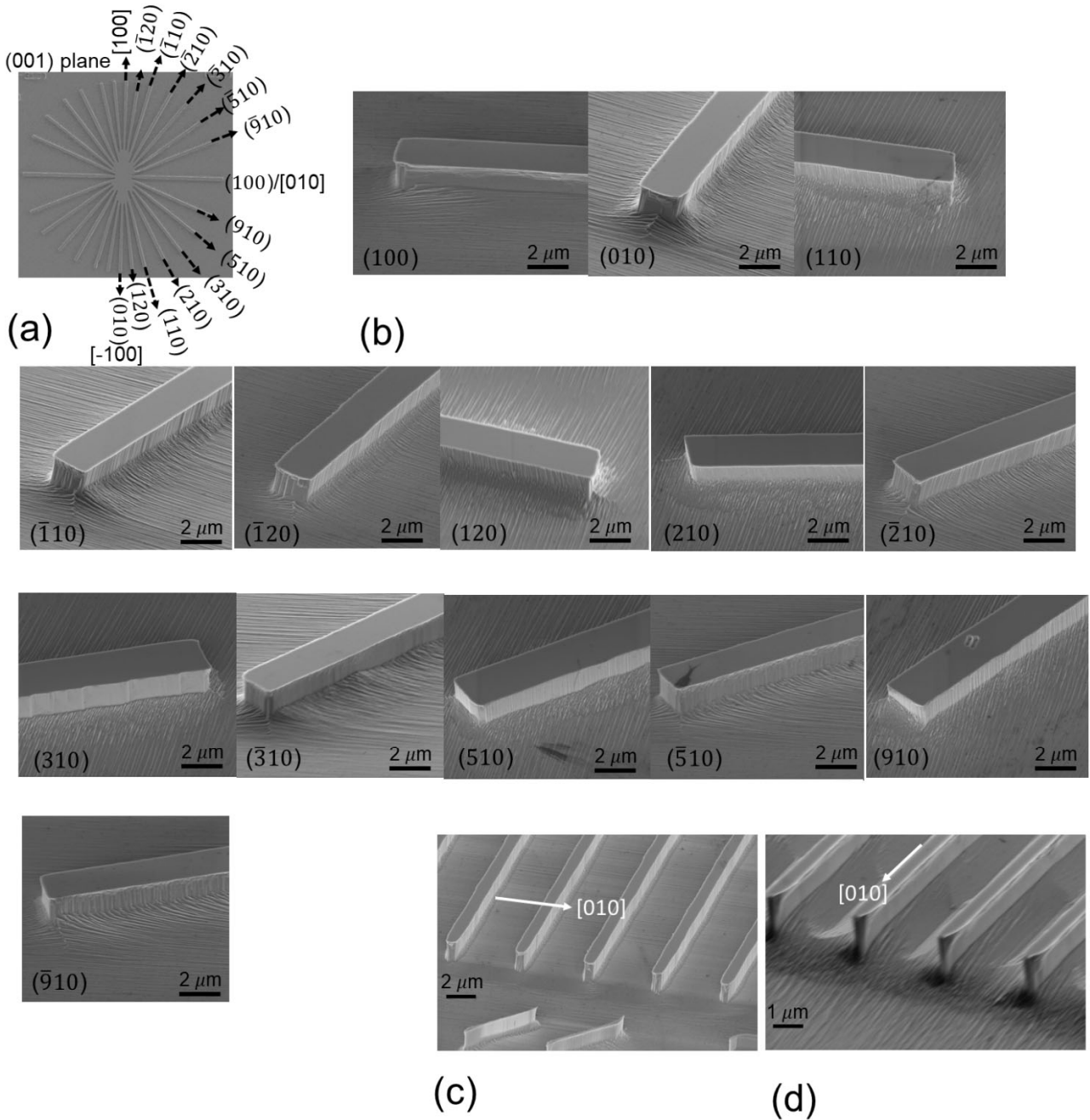

SFig.2 (a) Spoke wheel pattern used for etching study. Fins are marked according to the sidewall planes. (b) SEM images of fin sidewalls along different orientations. (c) Sub micron fins (900 nm wide) fabricated using Ga etching on (001) sample using Ga etch. (d) sub-micron fins (550 nm) fabricated on (001) sample using Ga etching.